%% file: main_topo_alloy.tex
\documentclass[aps,prl,twocolumn,floatfix,longbibliography,nofootinbib,superscriptaddress]{revtex4-2}
\usepackage[utf8]{inputenc}
\usepackage{natbib}
\usepackage{graphicx}
\usepackage{tikz}
\usepackage{xcolor}
\usepackage[tbtags]{amsmath}

\definecolor{brightmaroon}{rgb}{0.76, 0.13, 0.28}

\usepackage[colorlinks, linkcolor=blue]{hyperref}
\hypersetup{colorlinks,allcolors=brightmaroon}
\usepackage{orcidlink}
\usepackage{amssymb}
\usepackage{booktabs}
\usepackage{amsmath}
\usepackage{gensymb}
\usepackage{float}
\usepackage{amsmath}
\usepackage{tabularx,graphicx}
\usepackage{epstopdf}
\usepackage{latexsym}
\usepackage{color, colortbl}
\usepackage{psfrag}
\usepackage{bbm}
\usepackage{bm}
\usepackage{titlesec}
\usepackage{dsfont}
\usepackage{feynmp}
\usepackage{slashed}
\usepackage{multirow}
\usepackage[normalem]{ulem}
\renewcommand{\vec}[1]{\boldsymbol{#1}}

\newcommand{\bcen}{\begin{center}}
\newcommand{\ecen}{\end{center}}
\newcommand{\btab}{\begin{tabular}}
\newcommand{\etab}{\end{tabular}}
\newcommand{\bdes}{\begin{description}}
\newcommand{\edes}{\end{description}}

\newcommand{\beq}{\begin{equation}}
\newcommand{\eeq}{\end{equation}}
\newcommand{\bea}{\begin{eqnarray}}
\newcommand{\eea}{\end{eqnarray}}

\newcommand{\half}{\frac{1}{2}}
\newcommand{\bary}{\begin{array}}
\newcommand{\eary}{\end{array}}
\newcommand{\benum}{\begin{enumerate}}
\newcommand{\eenum}{\end{enumerate}}
\newcommand{\bitem}{\begin{itemize}}
\newcommand{\eitem}{\end{itemize}}

\newcommand{\bfig}{\begin{figure}}
\newcommand{\efig}{\end{figure}}
\renewcommand{\vec}[1]{\boldsymbol{#1}}
\newcommand{\bra}[1]{{\langle #1 |}}
\newcommand{\ket}[1]{| #1 \rangle}

\newcommand{\eqn}[1] {Eq.~(\ref{#1})}

\newcommand{\Fig}[1]{Fig.~\ref{#1}}

\makeatletter

\newcommand{\Rmnum}[1]{\expandafter\@slowromancap\romannumeral #1@}
\makeatother

\begin{document}

\title{Topological random alloy}


\author{Subrata Pachhal\orcidlink{0009-0008-2089-1759}}
\email{pachhal@iitk.ac.in}
\affiliation{Department of Physics, Indian Institute of Technology Kanpur, Kalyanpur, UP 208016, India}

\author{Aziz Hasan\orcidlink{0009-0007-0445-8237}}
\email{azizhasan2037@gmail.com}
\affiliation{Department of Physical Sciences, IISER Kolkata, Mohanpur, West Bengal 741246, India}

\author{Adhip Agarwala\orcidlink{0000-0003-1880-3429}}
\email{adhip@iitk.ac.in}
\affiliation{Department of Physics, Indian Institute of Technology Kanpur, Kalyanpur, UP 208016, India}

\begin{abstract} 
Topological phases of matter are often realized in crystalline materials. To extend their understanding beyond perfect stoichiometry, we introduce a minimal model of a topological random binary alloy and show that the system realizes an exotic form of impurity-band engineering. We reveal that, in contrast to Wannier charge centers pinned by impurities in conventional semiconductors, doping a proximate quantum anomalous Hall insulator results in dopant-centric chiral current loops. The nature of such current loops is intrinsically tied to the properties of both the host and the dopant. We demonstrate that, even at dilute dopant density, these current loops can form topological domains in an otherwise trivial host and trigger a topological phase transition. On the other hand, doping a topological host having chirality opposite to that of the dopants can unexpectedly stabilize a metallic phase in which bulk transport is mediated by inter-domain edge modes.
\end{abstract}

\maketitle

{\it Introduction.---}Impurity doping has long been a foundational tool for tailoring carrier concentrations and electronic properties in semiconductors \cite{sze_semiconductorbook_2021, grundmann_semiconductorbook_2006}. While semiconductors have {\it small} band gaps yet exhibit finite conductivity at room temperature due to thermally excited carriers, a class of insulators, described by symmetry-protected topology (SPT), supports robust conduction at their edges even at zero temperature \cite{Hasan_RMP_2010, Qi_RMP_2011, Shen_Book_2013, Bernevig_Book_2013}. Although the search for materials with such exotic character has traditionally relied on the bulk band structure of crystalline solids \cite{bernevig_nature_2017, bernevig_topmat_database}, several of them are realized in non-stoichiometric systems, including three-dimensional alloys such as  Bi$_x$Sb$_{1-x}$ and Pb$_x$Sn$_{1-x}$Te \cite{Hsieh_Bisb_2008, Xu_pbsnte_exp_2012, Yan_pbsnte_exp_20124}. In two dimensions, canonical examples are Cr- and V-doped (Bi,Sb)$_2$Te$_3$ thin films \cite{CIM_qah_2013, CIM_CrBST_2015, CIM_crbst2_2015} that exhibit the quantum anomalous Hall (QAH) effect, characterized by a quantized Hall response without any external magnetic field \cite{He_QAH_2013, Liu_QAH_2016}. Recently, Qu \textit{et al.}~\cite{Chan_topophotonicalloy_2024} demonstrated doping-driven topological phase transitions (TPTs) in photonic lattices, establishing a synthetic platform for exploring alloy physics. This stimulated a surge of experimental studies on photonic alloys \cite{Shi_highC_topo_photo_alloy_2024, Wang_quasicrystal_topo_photo_alloy_2024, Huang_amorphus_photo_topoalloy_2025, Wu_tpt_photo_alloy_2026}. Despite these achievements, theoretical studies of alloys in the context of topological phases have predominantly relied on the virtual crystal approximation and first-principles methods, ignoring their intrinsic disorder \cite{Chadov_VCAtopo_2013, Varjas_pbsnte_2020, CIM_bisb_2024}. While the interplay of disorder and topology is very rich both theoretically \cite{Li_TAI_2009, Groth_PRL_2009, Kobayashi_PRL_2013, Agarwala_PRL_2017, Ivaki_PRR_2020, Mondal_PRB_2023, mondal_IRFP_2025, Neehus_GTAI_2025, Zhang_topo_disorder_review_2026} and experimentally \cite{stutzer2018photonic, Liu_TAI_photonic_2020, LiPRL2021, zhou_amorphusTI_2020, Paul_amorphusbi2se3_2023, Zhang_amorphustopology_science_2023}, the effects of doping and the consequent formation of random alloys in SPT systems remain relatively underdeveloped.

\begin{figure}
    \centering
    \includegraphics[width=0.9\linewidth]{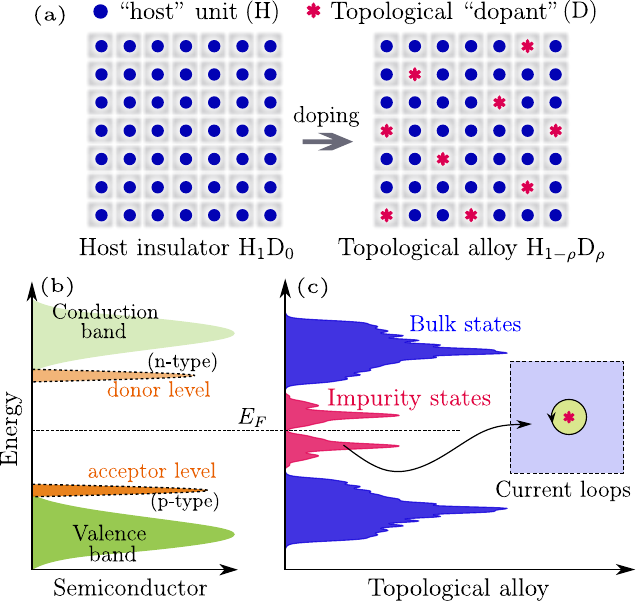}
    \caption{\textbf{Topological alloy:} (a) Doping in a pristine ``host" insulator (H$_{1}$D$_{0}$) by substituting the host {\it unit cells} (H) with topological ``dopants" (D) forms a topological alloy H$_{1-\rho}$D$_{\rho}$. (b) In doped semiconductors, either a donor level near the conduction band or an acceptor level near the valence band appears for n-type and p-type doping, respectively. (c) Topological doping creates impurity levels both above and below the Fermi energy $E_F$. These in-gap states can carry current loops around the impurity.}
    \label{fig1}
\end{figure}

In this work, we investigate the nature of impurity states induced by doping a proximate SPT phase---which we term \emph{topological doping}---and ask whether they imprint any nontrivial signatures in the system. To address these, we consider a two-dimensional host insulator and incorporate atomistic doping by randomly substituting {\it host units} (H) with topological {\it dopants} (D), which, upon complete substitution, form a pristine QAH insulator H$_{0}$D$_{1}$. At an intermediate dopant concentration $\rho$, the system therefore realizes a ``topological random alloy'', H$_{1-\rho}$D$_{\rho}$, as schematically illustrated in \Fig{fig1}(a). In conventional semiconductors, doping yields either donor levels (n-type) typically near the conduction band or acceptor levels (p-type) near the valence band (see \Fig{fig1}(b)). On the contrary, a topological dopant nucleates a pair of in-gap states, giving rise to impurity levels on both sides of the Fermi energy (see \Fig{fig1}(c)). We reveal that these in-gap states carry probability currents circulating around the impurity sites, inherited from their parent chiral edge states. Now, when the host insulator (H$_{1}$D$_{0}$) is trivial, one might expect that the dopant's topology will only take over in the alloy iff $\rho$ crosses the classical site percolation threshold \cite{Aharony_T&F_2003}. But remarkably, we find that impurity current loops in proximity hybridize via an effective tunneling and form topological domains that can render a QAH phase in an otherwise trivial host, even when $\rho$ is far below the classical threshold. Such TPT at strikingly low density can be seen in both Pb$_x$Sn$_{1-x}$Te \cite{Yan_pbsnte_exp_20124} and the photonic alloy \cite{Chan_topophotonicalloy_2024}. We also show that when hybridization is suppressed, the critical density approaches the classical value, indicating the influence of geometric connectivity.

In the scenario where the host H$_1$D$_0$ itself resides in a QAH phase with chirality opposite to that of H$_0$D$_1$, the above-mentioned TPT in the alloy (H$_{1-\rho}$D$_{\rho}$) is hindered. Instead, we find that at an intermediate $\rho$, the impurity-induced emergent domains coexist with the domains formed by the topological host, and their interface supports two co-propagating currents. This renders bulk conduction in the alloy and stabilizes a metallic phase. We demonstrate that, in contrast to the metallic phase in disordered topological superconductors \cite{Senthil_PRB_2000, Chalker_PRB_2001, Midenberger_PRB_2007, Medvedyeva_PRB_2010, Laumann_PRB_2012, Wang_PRB_2021} or random-bond Chern insulators \cite{Pachhal_RBCI_2025}, the edge-mediated metallicity in the topological random alloy is not restricted to particle-hole symmetric systems.

{\it Impurity doping on a QAH insulator.---}We consider a two-band insulator described by spinless fermion hopping between nearest-neighbor sites on a square lattice,
\begin{equation}
	H =  \sum_{i, \vec{\eta}}\Big( \Psi^\dagger_{i} T_{\vec{\eta}}\Psi_{i+\vec{\eta}} + \text{h.c.}  \Big)+ \sum_i \Psi^\dagger_i \Gamma_i \Psi_i,
	\label{eq_rham}
\end{equation}
where $\Psi_i = (c_{iA}~c_{iB} )^T$ and $\Psi^{\dagger}_i = (c^{\dagger}_{iA}~c^{\dagger}_{iB})$ with $c_{i\alpha}$, $c^{\dagger}_{i\alpha}$ representing annihilation and creation operators of orbital $\alpha \equiv (A, B)$ at $i^{\text{th}}$ site. Hopping along the direction $\vec{\eta}=(\hat{x}$, $\hat{y})$ is denoted by the matrices $T_{\vec{\eta}}=-\half\left(\sigma_{z}+i\sigma_{{\eta}}\right)$. The term $\Gamma_i=(2-M_i)\sigma_{z}$ accounts for the on-site energy, with Pauli matrices $\sigma_{x, y, z}$ acting in the orbital space. While the system breaks time-reversal symmetry, it follows $H\rightarrow H$ under particle-hole operation, $\Psi_i \rightarrow \sigma_x[\Psi_i^\dagger]^T, ~  \Psi^\dagger_i \rightarrow [\Psi_i]^T \sigma_x$, restricting it to symmetry class D of tenfold classification \cite{Altland_PRB_1997, Kitaev_AIP_2009, Ludwig_PS_2015}. 

With uniform {\it mass} term $M_i = M_h\forall i$, we model the undoped host (H$_1$D$_0$) which can be described by the Qi-Wu-Zhang (QWZ) model of Chern insulator: $\mathcal{H}_{\text{host}}(\vec{k}) = \sin k_x \sigma_x + \sin k_y \sigma_y +  \left(2-M_h-\cos k_x-\cos k_y\right)\sigma_z$ \cite{BHZ_Sci_2006, QWZ_spinlessBHZ_2006}. Thus, the host supports three topologically distinct phases, characterized by the Chern number $\mathcal{C} = \pm 1$ for $0 < M_h < 2$ and $2 < M_h < 4$ respectively, and $\mathcal{C} = 0$ for all other $M_h$. Insulator with $\mathcal{C} = \pm 1$ corresponds to the QAH phase with the anomalous Hall conductivity $\sigma_{xy} = \mathcal{C} e^2/ h$. Now we dope the system with concentration $\rho \in [0, 1]$ by randomly substituting host's (H) onsite energy $(2 - M_h)\sigma_z$ with dopant's (D) onsite energy $(2 - M_d)\sigma_z$. This can be achieved considering the following distribution of the mass term in \eqn{eq_rham},  
\begin{equation}
	\mathcal{P} (M_i) = (1-\rho)\delta(M_i - M_h) + \rho \delta(M_i - M_{d}).
	\label{eq_mdis}
\end{equation}
While the host H$_{1}$D$_{0}$ supports both trivial ($\sigma_{xy} = 0$) and topological ($\sigma_{xy} = \pm e^2/h$) phases tuned by $M_h$, we always stick to topological dopants with $M_d = 1$ such that H$_{0}$D$_{1}$ has only $\sigma_{xy} = e^2/h$ phase. The intermediate $\rho$ gives rise to a random binary alloy H$_{1-\rho}$D$_\rho$ via topological doping---a topological random alloy.

\begin{figure}
    \centering
    \includegraphics[width=1\linewidth]{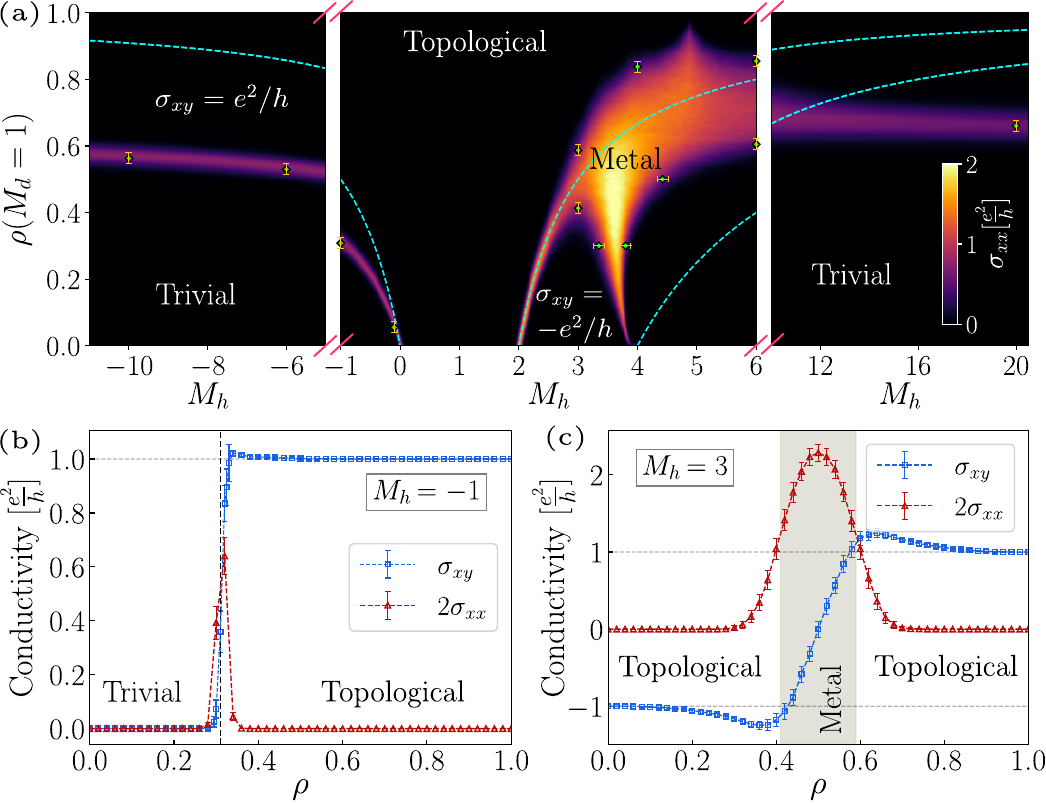}
    \caption{\textbf{Phase diagram:} (a) Configuration averaged $\sigma_{xx}$ phase diagram of the topological random alloy on a $60 \times 60$ lattice, computed over $420$ disorder realizations. Dashed lines are critical boundaries considering the virtual crystal approximation. The critical points marked with error bars are estimated via finite-size scaling. (b) Along $M_h = -1$, $\sigma_{xy}$ transition from the trivial and QAH phase, with a peak in $\sigma_{xx}$ near $\rho \approx 0.31$. (c) For $M_h = 3$, $\sigma_{xy} = \mp e^2/h$ phases are separated by a metallic region $0.41\lessapprox \rho \lessapprox0.59$ with large $\sigma_{xx}$. The data in (b) and (c) are averaged over $2000$ configurations.}
    \label{fig2}
\end{figure}

{\it Phases in topological random alloy.---}To analyze the features of the topological random alloy, we evaluate longitudinal and transverse conductivity $\sigma_{xx}$ and $\sigma_{xy}$ at $E_F = 0$ (half-filling) using \texttt{KWANT} \cite{Groth_Kwant_2014}. The details of the lead setup and calculation protocol are given in Supplemental Material (SM) \cite{sm}. Configuration averaged $\sigma_{xx}$ in the $M_h$-$\rho$ parameter space is illustrated in \Fig{fig2}(a). While at $\rho = 0$ the host with $M_h \leq 0$ is trivially insulating, the topological doping drives the alloy into a QAH phase at a certain density via a metallic critical point. For instance, \Fig{fig2}(b) shows at $M_h=-1$, $\sigma_{xy}$ jumps from $0$ to $e^2/h$ at $\rho_c \approx 0.31$ with peak in $\sigma_{xx}$ (see SM \cite{sm} for $\sigma_{xy}$ phase diagram). Moreover, we observe that when the host energy scale $M_h$ is comparatively small, say $M_h=-0.1$, even a small amount of dopants, as low as $\rho \sim 0.05$, can trigger the TPT in the alloy (see SM \cite{sm}). As $M_h$ becomes further negative, $\rho_c$ flows towards the classical site percolation threshold $\sim 0.6$ \cite{Aharony_T&F_2003}.

When the host itself is topological with the same chirality as the dopant's character (both $\mathcal{C} = +1$), i.e., $0<M_h<2$, expectedly the alloy will always remain in the $\sigma_{xy} = e^2/h$ QAH phase for all $\rho$. A more interesting feature emerges for $2<M_h<4$, where the pristine host (H$_{1}$D$_{0}$) is in $\mathcal{C} = - 1$ phase. With increasing $\rho$, the chirality of the alloy changes from host-type to the dopant-type via an extended region featuring macroscopic bulk conductivity instead of a conventional TPT. In \Fig{fig2}(c), we show the phases along $M_h = 3$, where between $\sigma_{xy} = \mp e^2/h$ insulators, the alloy stabilizes a metallic phase for $0.41 \lessapprox \rho \lessapprox 0.59$. Moreover, the metal is characterized by logarithmic growth of $\sigma_{xx}$ with the linear system size (see SM \cite{sm}). Although the metallic phase extends into the other trivial phase of the host $M_h>4$, as $M_h$ increases, the metal vanishes, leaving only the TPT in the alloy.

We estimate a few insulator-insulator and metal-insulator phase boundaries marked in \Fig{fig2}(a), using finite-size scaling of $\sigma_{xy}$ and $\sigma_{xx}$ (data in SM \cite{sm}). We also consider the virtual crystal approximation method, where an alloy is approximated by a perfect crystal \cite{Soven_CPA_PR1967}. In the topological random alloy, such an approximation is realized by replacing the mass distribution from \eqn{eq_mdis} with uniform mass: $(1-\rho)M_h + \rho M_d$ in the Hamiltonian \eqn{eq_rham} (see SM \cite{sm}). Following this, we obtain the {\it dashed} critical lines shown in \Fig{fig2}(a), which only predict insulating phases. This approach ignores intrinsic disorder in the topological alloy; thus, it is insufficient to explain the emergent metallic phase.

\begin{figure}
    \centering
    \includegraphics[width=1\linewidth]{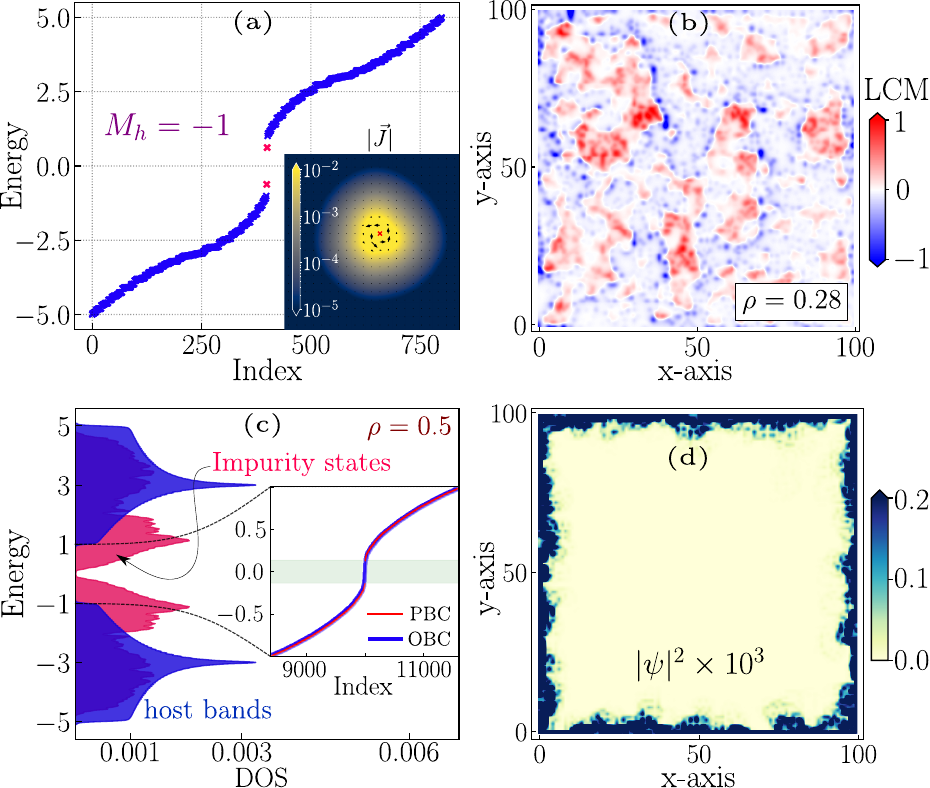}
    \caption{\textbf{Domain percolation landscape:} (a) Energy levels of an $L=20$ host with a single dopant and $M_h=-1$. The inset shows the chiral current of in-gap states around the cross-marked doped site. (b) LCM in a configuration of the alloy with $L=100$, $M_h = -1$, and $\rho = 0.28$ shows positively chiral domains. (c) While DOS at $\rho = 0.5$ with PBC shows gapped impurity states within the host bulk, (inset) zero-modes appear under open boundary conditions (OBC). (d) The wavefunction amplitude $|\psi|^2$ of these zero-modes lives on the global edge of the alloy.}
    \label{fig3}
\end{figure}

{\it Percolating current loops.---}In order to understand these transport features, let us first consider a single impurity doping on the host insulator. A topological dopant ($M_d = 1$) creates two impurity levels within the bulk gap of the trivial host, as shown in \Fig{fig3}(a) for $M_h = -1$. We evaluate the probability current vector, $\vec{J}_m = J_{m, x} \hat{x} + J_{m,y} \hat{y}$ induced by these in-gap states $\psi$ at each site ($m$) of the lattice where, $J_{m,x/y} =  i\big(\psi^{\dagger}_m T_{x/y} \psi_{m + \hat{x}/\hat{y}} - \psi^{\dagger}_{m + \hat{x}/\hat{y}} T^{\dagger}_{x, y} \psi_m\big)$ with $T_{x,y}$ are hoping matrices as defined below \eqn{eq_rham}. The inset of \Fig{fig3}(a) illustrates that $\vec{J}$ is distributed around the impurity like a loop having the same chirality as dopnat's parent topology. This can be thought of as a remnant edge state of a shrinking topological region \cite{Jha_current_CI_2017, Raquel_ringstate_2024} (see SM \cite{sm}). Now, when in proximity, these current-loop impurity states can hybridize with one another via the host. The effective hybridization Hamiltonian $H_{\text{eff.}}$ between two such current-loops can be derived by projecting the four eigenstates of the two-impurity problem onto their independent wavefunction basis (see SM \cite{sm}). Given two impurities $1$ and $2$, at a distance $r$ and angle $\theta$ with $\hat{x}$ we find,
\begin{align}
   H_{\text{eff.}}(r, \theta) = \begin{pmatrix}
       H_{11} & H_{12}\\
       H^{\dagger}_{12} & H_{22}
   \end{pmatrix},
\end{align}
where $H_{11} =H_{22} \approx 0.68 \sigma_z$ acts like onsite mass term and $H_{12} \approx 0.36 e^{-0.4r} \sigma_z - 0.19 i e^{-0.6r} \left(\cos \theta \sigma_x + \sin \theta \sigma_y\right)$ mimic a hoping $T(r, \theta)$ similar to a generalized QWZ model \cite{Agarwala_PRL_2017}. Such hybridization allows the current loops to form bigger loops and, given the effective mass $M_{\text{eff}}\approx (2-0.68) = 1.32$, they eventually become topological domains even at a very small doping density. The domain formation can be visualized via the ``local Chern marker" defined at each site by, $\text{LCM} (x, y) = -2\pi \text{Im} \big(\text{Tr}_{\text{site}}\left[PxP,PyP\right]\big)$, where $P = \sum_{n=1}^{N/2}\ket{\psi_n}\bra{\psi_n}$ is the many body ground state Projector at half-filling \cite{Resta_LCM_2011}. In \Fig{fig3}(b), we show in a representative configuration for $M_h = -1$, the dopants at density $\rho =0.28$ give rise to emergent topological domains with positive LCM. The negative LCM regions represent the domain boundary with the trivial host, where the chiral current flows. Now, as $\rho$ increases, topological domains grow and merge with each other (see SM \cite{sm}), similar to the percolation landscape of the plateau transition in integer quantum Hall systems \cite{Chalker_percoIQHE_1988}. This drives a TPT in the alloy even before the dopants can proliferate classically, which occurs only after $\rho \simeq 0.6$ in a square lattice \cite{Aharony_T&F_2003}. For instance, at $\rho=0.5$, the DOS in \Fig{fig3}(c) shows that the gap of the impurity states already hosts zero-modes in open boundary conditions, and those zero-modes form a global edge state (see their wavefunction amplitude $|\psi|^2$ in \Fig{fig3}(d)), indicating a QAH phase in the alloy.

The domain percolation process is dominant when the gap scale $M_h$ of the trivial host is comparable to $M_d = 1$, such that the quantum tunneling between current loops is finite. However, as $M_h$ becomes increasingly negative, tunneling is suppressed as $\sim |M_h|^{-1}$ (see SM \cite{sm}), hindering domain formation. Thus, as $M_h \rightarrow -\infty$, a global edge state can only form after the dopants go through a classical percolation transition, and the critical density for TPT in the alloy flows towards $\rho \rightarrow 0.6$. 
 
\begin{figure}
    \centering
    \includegraphics[width=1\linewidth]{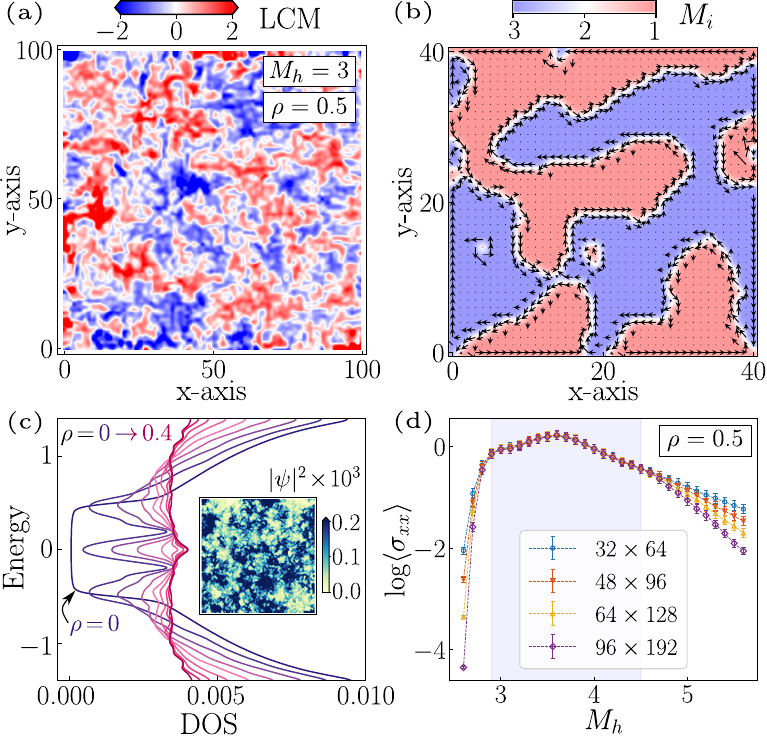}
    \caption{\textbf{Edge mediated metal:} (a) LCM in a snapshot of the $L=100$ alloy with topological host $M_h =3$ and $\rho=0.5$. (b) In a $L=41$ lattice, correlated patches of $M_i = 1$ and $3$, both positive and negative LCM domains arise with co-propagating currents at their boundaries. (c) For $M_h=3$, increasing $\rho$ enhances the low-energy DOS; the inset illustrates that these states (parameters the same as (a)) are extended in the lattice. (d) With Anderson disorder $U =1$ and particle-hole symmetry breaking term $\gamma = 0.1$, $\sigma_{xx}$ (averaged over $1400$ configurations) vs $M_h$ at $\rho=0.5$ exhibits a metallic phase in a class A alloy with scale-invariant conductivity.}
    \label{fig4}
\end{figure}

{\it Emergence of a metal.---}Domain formation driven by current loops also occurs when the host itself is topological with opposite chirality, i.e., $2 < M_h < 4$. However, in this scenario, since the current loops carry the information of the host's topology (see SM \cite{sm}), domains of both chirality---one from the dopants, another from the host---coexist in the alloy. 
\Fig{fig4}(a) illustrates one such configuration for $M_h = 3$ and $\rho = 0.5$, where domains with positive and negative LCM emerge in the alloy. This allows for co-propagating chiral edge states at their boundaries, leading to the formation of {\it snake}-like states \cite{Lambert_graphenesnake_2008}. To illustrate this, we generate larger domains by artificially introducing correlations among dopants (see SM \cite{sm}), and show the current flow through the boundaries of both type domains in \Fig{fig4}(b), for the same parameter values as in \Fig{fig4}(a). 

At densities where domains induced by both host and dopant cover comparable regions, an abundance of such snake-like states (similar to \Fig{fig4}(b)) can create a finite density of mid-gap states. This is confirmed in \Fig{fig4}(c), where we show that the DOS increases near zero energy as $\rho$ increases to 0.4. These states are extended in the lattice (inset of \Fig{fig4}(c)), and allow bulk transport leading to metallicity in the alloy with $\sigma_{xx} \sim \log(L)$ as discussed near \Fig{fig2}(c). Since such transport occurs via inter-domain scattering of snake states, trivial domains in a topological host will just pinch off conduction channels. We demonstrate this by studying transmission around a single domain, given the topology of the host, and dopants are varied (see SM \cite{sm}).

While in disordered topological superconductors, a thermal metal arises via proliferation of Majorana bound states \cite{Senthil_PRB_2000, Chalker_PRB_2001, Midenberger_PRB_2007, Medvedyeva_PRB_2010, Laumann_PRB_2012, Wang_PRB_2021}, an electronic metal emerges in a random bond-disordered Chern insulator due to percolation of zero-modes nucleated through $\pi$-flux \cite{Pachhal_RBCI_2025}; in both cases, particle-hole symmetry is crucial. Since the metallic phase of the topological random alloy is mediated by the chiral edge modes, it is not subject to symmetry restrictions. To demonstrate this, we perturb the alloy Hamiltonian in \eqn{eq_rham} using,
\begin{align}
    H_{\text{A}} = H + \frac{\gamma}{2} \sum_{i, \vec{\eta}} \big(\Psi^\dagger_{i} \Psi_{i+\vec{\eta}} + \text{h.c.}\big) + \sum_i \Psi^\dagger_i U_i \Psi_i,
\end{align}
where both the hoping $\gamma$ and the Anderson disorder $U_i \in [-U/2, U/2]$ breaks particle-hole symmetry. In this class A alloy with $\gamma = 0.1$, and $U = 1$, following the same random mass distribution as in \eqn{eq_mdis}, we present how the $\sigma_{xx}$ varies with $M_h$ in \Fig{fig4}(d), for $\rho = 0.5$. The region $2.9 \lessapprox M_h \lessapprox 4.5$ shows scale-invariant conductivity signaling a critical metal phase with no $\sigma_{xx}\sim \log(L)$ correction compared to $\gamma=0$ (see SM \cite{sm}). While Anderson localization is generically expected in class A \cite{Evers_localizationreview_2008}, we wonder whether the critical metal here arises from an underlying disordered Dirac fermion with a perturbative mass, which can evade localization \cite{Zhang_metalmassiveDirac_2021, Ostrovsky_graphenemetal_2007, Bardarson_graphenemetal_2007, Nomura_metalDirac_2007}.

{\it Outlook.---}With the growing ability to engineer topological phases in both materials and synthetic platforms, disorder introduced by doping and alloying can become an important route to realizing and controlling SPT \cite{Chan_topophotonicalloy_2024, Shi_highC_topo_photo_alloy_2024, Wang_quasicrystal_topo_photo_alloy_2024, Huang_amorphus_photo_topoalloy_2025, Wu_tpt_photo_alloy_2026}. In this work, we introduce a minimal model of
a topological random alloy. We show that topological doping in a host insulator gives rise to chiral current loops originating from both the host and the dopant. Non-trivial domains formed by such current loops can endow an otherwise trivial host with topological character even at very low doping densities. Remarkably, when the host itself is topological and has chirality opposite to that of the dopant's parent topology, a metal can emerge in the alloy mediated by conducting interfaces between domains of competing chirality. This metal is shown to be stable even in the absence of intrinsic symmetries. Our work unifies the role of geometrical connectivity and quantum correlations in realizing topological random alloys.

While our work is designed to create a minimal model, first-principles studies predicting such alloys would be a welcome step. Moreover, exploring the effects of dopants with local magnetic moments can have important implications in understanding Kondo-like physics \cite{Anderson_kondo_1961} and spin-glassy behavior in such doped topological insulators.

{\it Acknowledgment.---} We thank Koushik Pal, Soumya Bera, Naba P. Nayak, and Soumya Sur for insightful comments and discussions. S.P. acknowledges funding from the IITK Institute Fellowship. A.H. acknowledges funding from the UGC. Numerical calculations were performed on the workstations \texttt{WIGNER} and \texttt{SYAHI} at IITK.

\bibliography{perco_ref}

\include{SM_topo_alloy_add_to_main.tex}

\end{document}

%% file: SM_topo_alloy_add_to_main.tex
\newpage
\setcounter{equation}{0}
\setcounter{table}{0}
\setcounter{figure}{0}
\makeatletter
\renewcommand{\thesection}{S\Roman{section}}
\renewcommand{\theequation}{S\arabic{equation}}
\renewcommand{\thefigure}{S\arabic{figure}}
\renewcommand{\thetable}{S\arabic{table}}
\onecolumngrid

\begin{center}
	\textbf{\large Supplemental Material to ``Topological random alloy"}\\
    \vspace{0.4cm}
    {Subrata Pachhal$^1$, Aziz Hasan$^{2}$, Adhip Agarwala$^1$}
    \\
    \vspace{0.2cm}
    \small{\it $^1$Department of Physics, Indian Institute of Technology Kanpur, Kalyanpur, UP 208016, India}\\
    \small{\it $^2$Department of Physical Sciences, IISER Kolkata, Mohanpur, West Bengal 741246, India}

\end{center}

\vspace{\columnsep}

\twocolumngrid

\section{SI.~Conductivity calculation}
\begin{figure}
    \centering
    \includegraphics[width=0.7\linewidth]{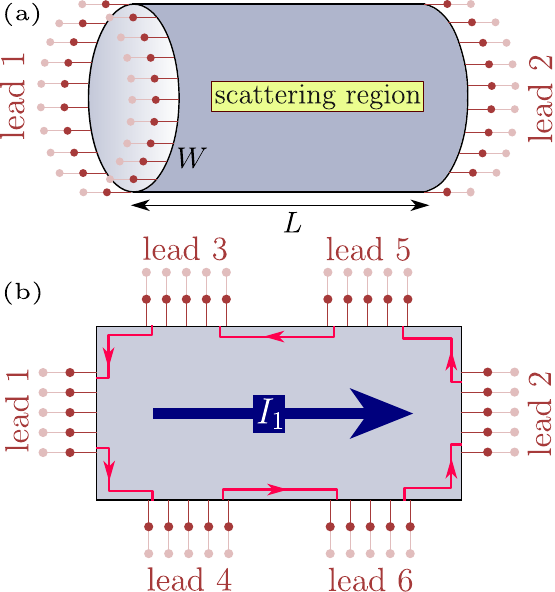}
    \caption{\textbf{\texttt{KWANT} setup:} (a) Two-terminal setup where a scattering region with length $L$ and width $W$ is cylindrical. Two leads, composed of one-dimensional wires, are connected to the left and right edges. (b) Six-terminal geometry where we always choose $L = 2W$, such that all the $6$ leads are uniformly placed. Here, $I_1$ is the source current.}
    \label{smfig1}
\end{figure}

We employ \texttt{KWANT} \cite{Groth_Kwant_2014} to calculate longitudinal and transverse conductivity $\sigma_{xx}$ and $\sigma_{xy}$. For $\sigma_{xx}$, a two-terminal setup is used with periodic boundary conditions (PBC) in the transverse direction, making it a cylindrical geometry as shown in \Fig{smfig1}(a). The two leads connected to the left and right edges are modeled as a set of semi-infinite one-dimensional metallic wires with Hamiltonian $H_{\text{lead}} = -\sum_i \left(\Psi_{i}^{\dagger} \Psi_{i+1} + h.c.\right)$. Since we are working with a two-band system, $\Psi$ in the lead is also a two-component object as described in the main text. \texttt{KWANT} generates the transmission probability $T_{12}$ between lead 1 and lead 2; thus, the conductance is $G_{12} = (e^2/h)T_{12}$ following the Landauer-Buttiker formula \cite{Datta_Book_1997}. Given the length and width of the scattering region are $L$ and $W$ respectively, the longitudinal conductivity is, 
\begin{equation}
    \sigma_{xx} = \frac{L}{W} G_{12} = \frac{L}{W}T_{12}\frac{e^2}{h}.
\end{equation}

For the $\sigma_{xy}$ calculation, we use a six-terminal device as shown in \Fig{smfig1}(b) and consider $L = 2W$ with uniformly placed leads to reduce the geometric effects. Calculating the transmission $T_{mn}$ between lead $m$ and $n$ using \texttt{KWANT} (note $T_{mm} = 0$), we get the current at different leads from the following relation,
\begin{equation}
I_m =  \frac{e^2}{h} \sum_{n\neq m}\big(T_{nm}V_m - T_{mn}V_n\big),\label{eq_sixterminal} 
\end{equation}
where $m$ and $n$ run from 1 to 6 and $V_{m/n}$ is voltage at lead $m/n$. While leads 1 and 2 serve as the current input, leads 3, 4, 5, and 6 serve as voltage probes. To force these conditions, we fix $V_1 = 1, V2 = 0$ and $I_{3/4/5/6} = 0$ such that $I_1 = -I_2$ and $V_{3/4/5/6}$ are the unknown voltages. Solving \eqn{eq_sixterminal}, we get these unknown quantities and calculate longitudinal and transverse resistances as follows, 
\begin{align}
R_{xx} = \half \left( \frac{V_3 - V_5}{I_1} + \frac{V_4 - V_6}{I_1} \right), \\
R_{xy} = \half \left( \frac{V_3 - V_4}{I_1} + \frac{V_5 - V_6}{I_1} \right).
\end{align}
Using these we form the resistance tensor with $R_{yy} = R_{xx}$ and $R_{yx} = - R_{xy}$. The averaging further reduces the effects of anisotropic geometry. Finally, inverting the resistance tensor, we get the Hall conductivity $\sigma_{xy}$ in units of $\frac{e^2}{h}$. 

\section{SII.~Virtual crystal approximation}
\begin{figure}
    \centering
    \includegraphics[width=1\linewidth]{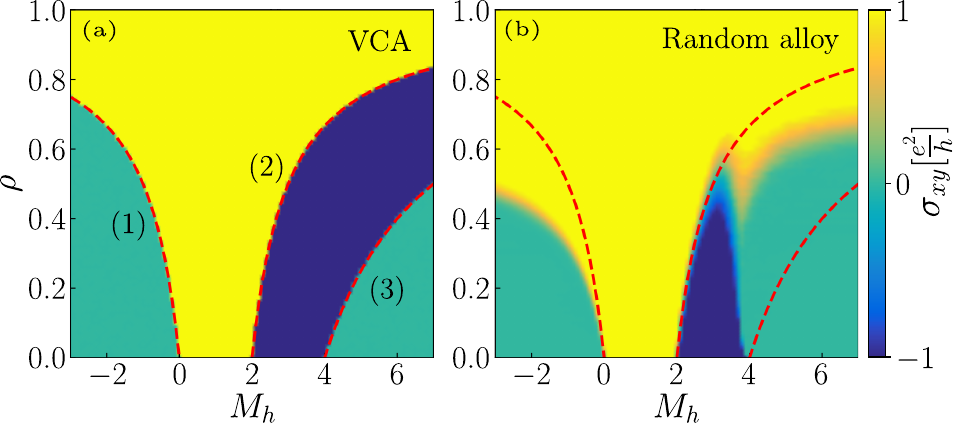}
    \caption{\textbf{Phase diagram} (a) $\sigma_{xy}$ phase diagram for the topological alloy under virtual crystal approximation (VCA) (see \eqn{eq_mfham} with phase boundaries denoted by dashed lines. (b) Phase diagram for the topological random alloy with intrinsic disorder. $\sigma_{xy}$ is averaged over $420$ configurations. For both (a) and (b), system sizes are $L = 80$ and $W = L/2 = 40$, and we always consider $M_d=1$.}
    \label{smfig2}
\end{figure}

Generally, in conventional studies of alloys, the non-uniform potentials of different atoms are replaced by an average uniform potential landscape using the virtual crystal approximation (VCA), also known as the coherent potential approximation \cite{Soven_CPA_PR1967}. We use this approximation in our system (Hamiltonian given by Eq. (1) in the main text), where instead of the random mass distribution $\mathcal{P} (M_i) = (1-\rho)\delta(M_i - M_h) + \rho \delta(M_i - M_{d})$ in the actual binary topological alloy, we use a uniform mass: $\tilde{M} = (1-\rho) M_h + \rho M_{d}$ and restore the translational invariance. The approximated Hamiltonian in the momentum space is given by the QWZ model \cite{QWZ_spinlessBHZ_2006}, 
\begin{align}
	\mathcal{\tilde{H}}_{\text{alloy}}(\vec{k}) & = \sin k_x \sigma_x + \sin k_y \sigma_y\nonumber \\ 
    & ~~~~+ \big(2-\tilde{M}-\cos k_x-\cos k_y\big)\sigma_z.
    \label{eq_mfham}
\end{align}
with re-normalized mass $\tilde{M}$. Thus, $\mathcal{\tilde{H}}_{\text{alloy}}$ has two QAH phases with $\sigma_{xy} = \pm e^2/h$ for $0<\tilde{M}<2$ and $2<\tilde{M}<4$ respectively, and they are separated by critical points at $\tilde{M} =0, 2$ and $4$. In terms of the constituent masses of the alloy, this gives three critical densities: (1) $\rho = M_h /(M_h-M_d)$, (2) $\rho = (M_h-2) /(M_h-M_d)$, and (3) $\rho = (M_h - 4) /(M_h-M_d)$. Considering $M_d=1$, in \Fig{smfig2}(a), we show the $\sigma_{xy}$ phase diagram in $M_h$-$\rho$ parameter space of the topological random alloy under VCA (see \eqn{eq_mfham}) along with the three critical lines. For the topological alloy with intrinsic randomness in the mass distribution, the configuration-averaged $\sigma_{xy}$ phase diagram is shown in \Fig{smfig2}(b).

\section{SIII.~Features of the metallic phase}
\begin{figure}
    \centering
    \includegraphics[width=1\linewidth]{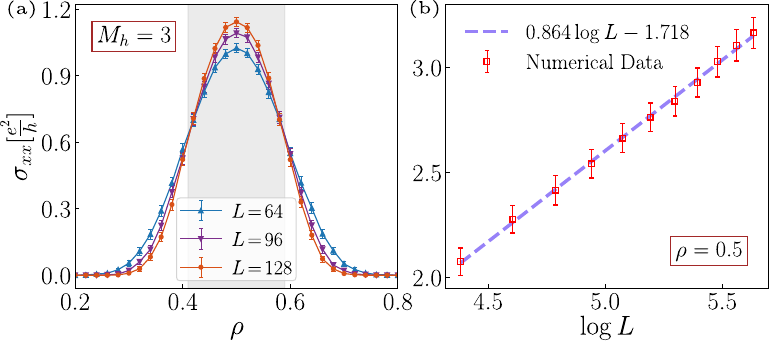}
    \caption{\textbf{Metallic features:} (a) Along $M_h = 3$, $\sigma_{xx}$ with $\rho$ for three different system sizes $L = 64, 96, 128$. Within the shaded region, $\sigma_{xx}$ increases with $L$. (b) The increment in $\sigma_{xx}$ show $\log L$ signature for $\rho=0.5$. In both (a) and (b), a $L/2 \times L$ setup is used, such that the aspect ratio is always $2$, and $\sigma_{xx}$ data are averaged over $1400$ configurations.}
    \label{smfig3}
\end{figure}

To see the metallic feature of the topological random alloy, we consider $M_h = 3$ and show $\sigma_{xx}$ as a function of $\rho$ for three system sizes in \Fig{smfig3}(a). While the conductivity for $0.41 \lessapprox  M_h \lessapprox 0.59$ increases with linear system size L, indicating the metallic phase, for all other $\rho$, $\sigma_{xx}$ is suppressed, showing insulating behavior. In the metallic phase, $\sigma_{xx}$  increases logarithmically with increasing $L$, as shown in \Fig{smfig3}(b) for $\rho =0.5$.

\begin{figure}
    \centering
    \includegraphics[width=1\linewidth]{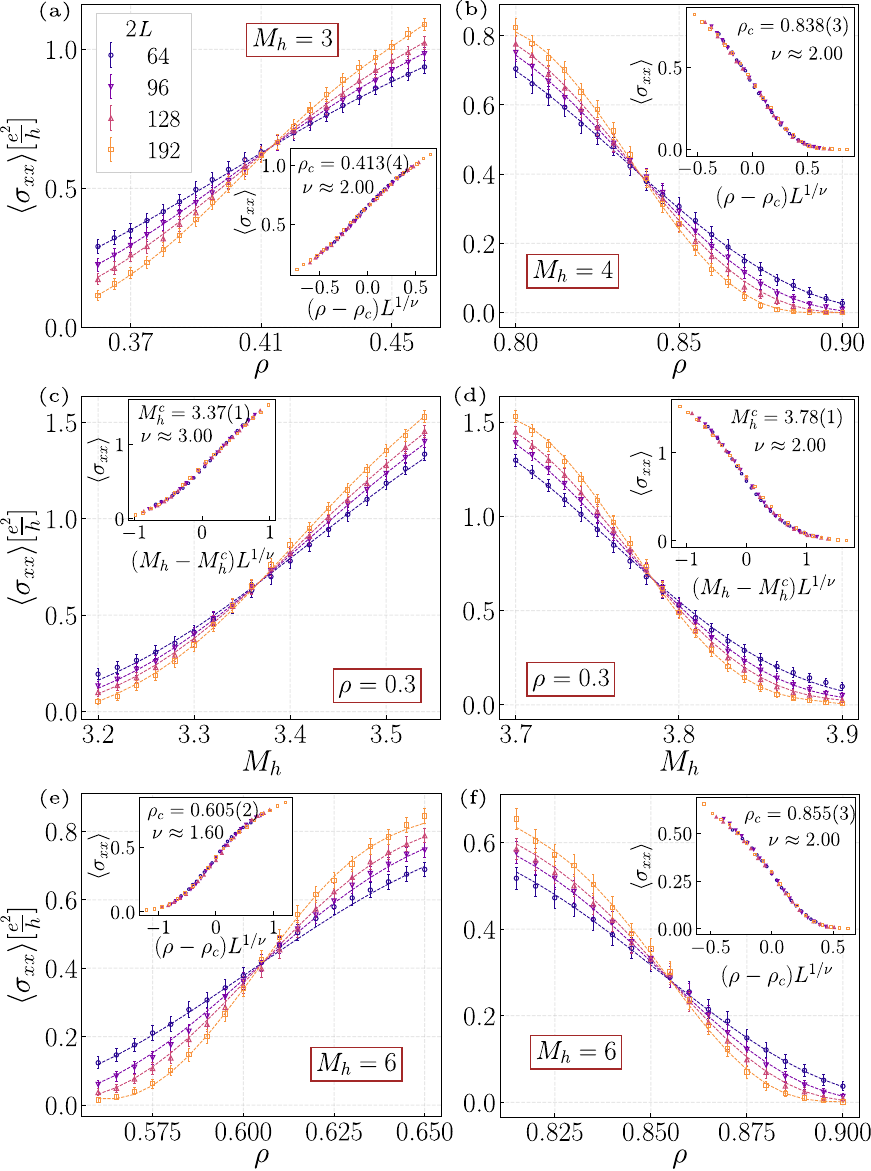}
    \caption{\textbf{Metal-insulator transitions:} (a-f) Metal-insulator critical points for various parameter values are estimated using system size scaling of configuration-averaged $\sigma_{xx}$ calculated in a $L\times 2L$ set up. Their insets show corresponding scaling collapses with estimated critical parameter values. Number of configurations used are $\mathcal{N}_c = 2800$ for $2L=64, 96, 128$ and $\mathcal{N}_c = 1400$ for $2L=192$.}
    \label{smfig4}
\end{figure}

\section{SIV.~Finite size scaling of conductivity}
In this section, we outline the fitting procedure for estimating the critical parameters associated with the phase boundaries of the topological random alloy. In general, observables are expanded in terms of both relevant and irrelevant scaling variables up to a given order, and the resulting scaling function is then fitted to numerical data to perform finite-size scaling analysis. The expansion of the order parameter (here, disorder-averaged conductivity $\langle{\sigma_{xx/xy}} \rangle$) is given as follows,
\begin{equation}
\langle{\sigma_{xx/xy}} \rangle = f_0(x) + b_0 L^{-y} f_1(x) 
\label{eq_g_expand}
\end{equation} 
where, $f_j(x) = \sum_{n=0}^{N_\text{R}} a_{j n} x^n$, $x = (z-z_\text{c})/z_\text{c} \cdot  L^{1/\nu}$, $z$ being the parameter that drives the transition at $z_c$. $\nu$ and $y$ are the leading relevant and irrelevant exponents, respectively; $b_0$ and $a_{jn}$ are expansion coefficients, with $N_R$ being the order of relevant expansion.

\begin{figure*}
    \centering
    \includegraphics[width=1\linewidth]{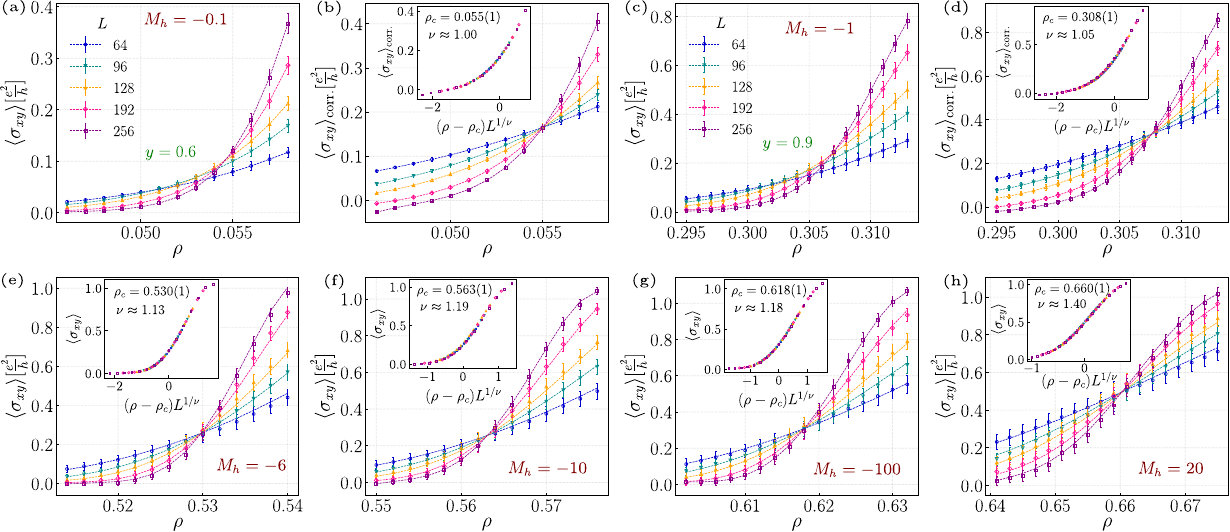}
    \caption{\textbf{Insulator-insulator transitions:} Insulator-insulator critical points are estimated using system size scaling of configuration-averaged $\sigma_{xy}$. Number of configuration used for $L=64, 96, 128$ is $\mathcal{N}_c = 2800$, for $L=192$, $\mathcal{N}_c = 1400$ and for $L=256$, $\mathcal{N}_c = 700$. (a) Raw data for $M_h=-0.1$ show a shift in crossing with consecutive $L$, indicating a strong finite size correction with exponent $y=0.6$. (b) Corrected data and the scaling collapse with critical point $\rho_c = 0.055 \pm 0.001$ in the inset. (c,d) Same as (a) and (b) but for $M_h = -1$ and the correction is $y=0.9$. (e-h) Scaling data for $M_h = -6, -10, -100, 20$ with fitted curve. Their corresponding insets show scaling collapse with respective $\rho_c$.}
    \label{smfig5}
\end{figure*}

To see the metal-insulator transitions, using a $L\times 2L$ lattice (such that the aspect ratio is $2$), we evaluate $\langle \sigma_{xx} \rangle$ as a function of the driving parameter ($z = M_h$ or $\rho$) near six of the speculated metal-insulator phase boundaries (see phase diagram Fig.~2(a) in the main text) for $4$ different system sizes as shown in \Fig{smfig4}(a-f). With $N_R = 5$, the function in \eqn{eq_g_expand} is fitted to the data using a non-linear least-squares approach. Note that the irrelevant exponent $y$ is large for metal–insulator critical points and can thus be neglected. The corresponding fitted functions with their estimated critical parameters $z_c= M_h^{c}$ or $\rho_c$ along with the scaling collapses are shown in the inset of \Fig{smfig4}(a-f).

Similarly, for six different $M_h$ near the insulator-insulator phase boundaries (see phase diagram Fig.~2(a) in the main text), we evaluate $\langle \sigma_{xy} \rangle$ as a function of $\rho$ using a $L \times L/2$ lattice for $5$ different values of L as the data are shown in \Fig{smfig5}(a-h). Here, the function in \eqn{eq_g_expand} is fitted with $N_R = 3$ to avoid overfitting, and the exponent $y$ is neglected except for $M_h=-0.1$ (\Fig{smfig5}(a)) and $M_h=-1$ (\Fig{smfig5}(c). For these two parameters, the corrected data $\langle \sigma_{xy} \rangle_{\text{corr.}} = \langle \sigma_{xy} \rangle - b_0 L^{-y} f_1(x)$, are shown in \Fig{smfig5}(b) and (d), respectively, and their collapse with respective critical points, $\rho_c$, are shown in the corresponding insets. 

We note that the critical exponent for metal-insulator transitions varies strongly around $\nu \approx 2$, and for insulator-insulator transitions it's around $\nu \approx 1$, with large error bars as per our data. Thus, establishing the universality classes of these transitions in the topological random alloy is beyond the scope of this article and remains a challenge for the future.

\begin{figure}
    \centering
    \includegraphics[width=1\linewidth]{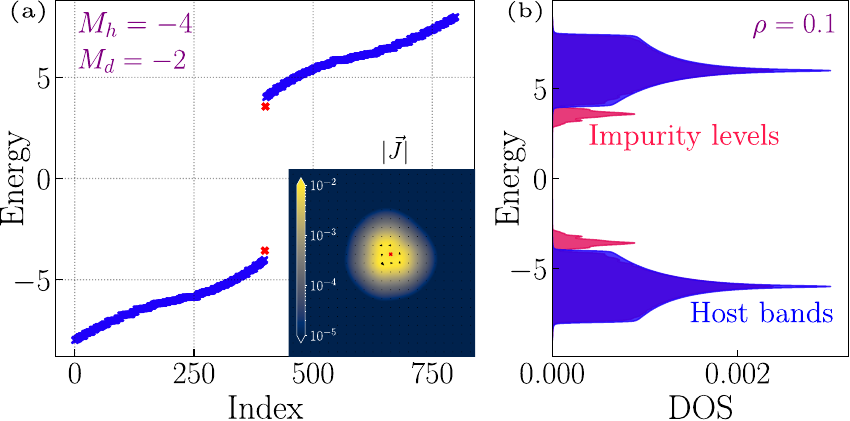}
    \caption{\textbf{Trivial doping:} (a) Energy levels of an $L=20$ host with a single dopant for $M_h=-4$ and $M_d = -2$. The inset shows the current profile of in-gap states. (b) With the same $M_h$ and $M_d$, the DOS at $\rho = 0.1$ shows impurity levels appear near the valence and conduction band of the host.}
    \label{smfig6}
\end{figure}

\begin{figure*}
    \centering
    \includegraphics[width=0.85\linewidth]{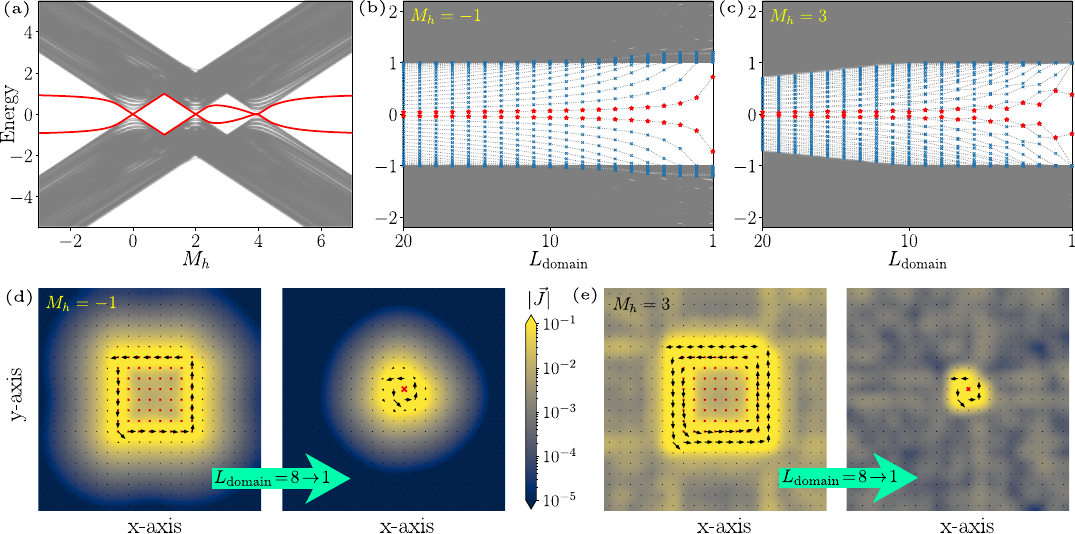}
    \caption{\textbf{Origin of current loop:} (a) Energy spectrum of the host in the presence of a single impurity doping with $M_d=1$ and $L=30$. Two impurity levels are shown in red. (b) We consider a domain of linear size $L_{\text{domain}}$ with $M_d=1$ inside a trivial host with $M_h = -1$ and $L=30$. The energy levels show that only two modes remain within the bulk gap as the domain shrinks to a single impurity at $L_{\text{domain}} = 1$. (c) Same as (b) but for the topological host with $M_h = 3$. (d) For $M_h=-1$, the current profile of the in-gap modes shows a chiral edge state in the domain boundary for $L_{\text{domain}} = 8$, and a remnant of it can be observed even for $L_{\text{domain}} = 1$. The red crosses are the domain sites. (e) For $M_h=3$, the domain boundary and the boundary of the host both support co-propagating edge-states, giving rise to broader current distribution in $L_{\text{domain}} = 1$ limit.}
    \label{smfig7}
\end{figure*}

\section{SV.~Current-loops of impurity levels}
To illustrate the current loops, we consider a single impurity in the host insulator. In \Fig{smfig6}(a), we show the two impurity levels for a trivial host with $M_h=-4$ and a single trivial impurity with $M_d = -2$. The current distribution in the lattice for these two in-gap states, as shown in the inset of \Fig{smfig6}(a), is very faint, compared to the case of a topological dopant in a trivial host (see Fig.~3(a) in the main text). At a finite density of such trivial doping, the impurity levels appear only near the valence and conduction bands of the host, mimicking the acceptor and donor levels of the semiconductor \cite{grundmann_semiconductorbook_2006}, but both together. The DOS in \Fig{smfig6}(b) represents such a situation for $M_h=-4, M_d=-2$ at $\rho = 0.1$.

A topological dopant gives rise to impurity states within the bulk gap if the host is either trivial or topological with opposite chirality compared to the dopant's parent topology, as shown in \Fig{smfig7}(a) for $M_d = 1$. In the main text, we show that the in-gap states exhibit a prominent chiral current loop centered on the doped site for $M_h = - 1$ (Fig.~3(a)). To understand the origin of such exotic impurity states, we consider a domain of size $L_{\text{domain}} \times L_{\text{domain}}$ with $M_d=1$ in the center of the host insulator, such that as $L_{\text{domain}} \rightarrow 1$, it represents a single impurity doping. The energy spectrum in \Fig{smfig7}(b) shows, for the trivial host ($M_h=-1$), with shrinking $L_{\text{domain}}$, the in-gap modes associated with the edge state of the domain, mix with the bulk; only two remain in the single impurity limit. A similar feature can be seen in \Fig{smfig7}(c) for the topological host with $M_h=3$. In \Fig{smfig7}(d), for $M_h = -1$, we illustrate how the current profile of the in-gap modes changes from the chiral edge state at the domain boundary to the loop around the impurity. For $M_h=3$, the boundary of the host adjacent to the domain also supports an edge state, thus two co-propagating current flows as shown in \Fig{smfig7}(e) and give rise to the current loop tied to the nature of both the host and the impurity.

\begin{figure*}
    \centering
    \includegraphics[width=1\linewidth]{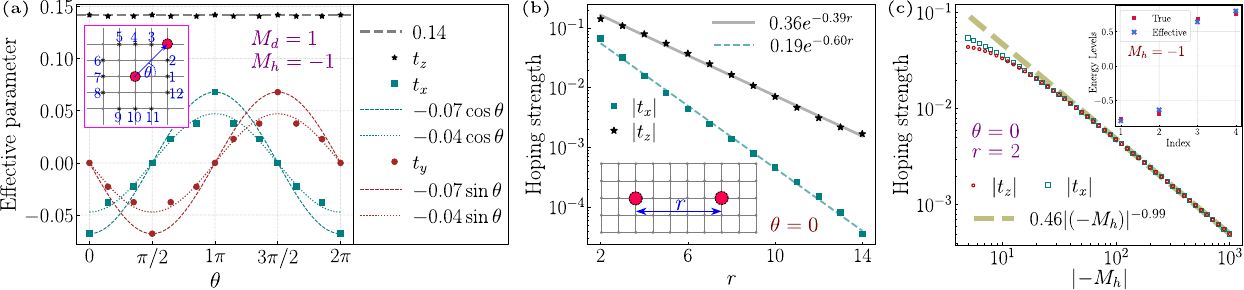}
    \caption{\textbf{Effective hopping structure:} (a) For $M_h=-1$ and $M_d=1$, hopping parameter $t_{x/y/z}$ (component of Pauli matrices $\sigma_{x/y/z}$ in the effective Hamiltonian) between two impurities as a function of $\theta$, where $\theta$ is the angle as shown in the inset. As we go $1, 2, 3,..., 12, 1$, $\theta$ changes from $0\rightarrow 2\pi$ but the distance $r\approx 2$ changes slightly. Also, note that $t_x$ and $t_y$ have the same amplitude. (b) Along $\theta = 0$, the strength hoping parameter $t_x$ and $t_z$ ($t_y = 0$) as a function of $r$. (c) With $M_h$ going far negative, the hoping strength decreases rapidly as shown for $\theta=0$ and $r=2$. The inset shows the in-gap modes of the two-impurity problem and the energy levels of the effective Hamiltonian. For all the calculations, $L=30$.}
    \label{smfig8}
\end{figure*}

\section{SVI.~Tunneling between two impurities}
In this section, we show that the current loops associated with impurity levels can tunnel between each other via an effective hybridization when in proximity. Since these impurity states are not just charge centers, we can not Wannierize them to evaluate the hybridization. Instead, we use the following protocol:
\begin{enumerate}
    \item Given a host insulator, evaluate two in-gap states for a single impurity at position $\vec{r_1}$: $\ket{\alpha_1}$ and $ \ket{\beta_1}$.
    \item Independently solve for another impurity at $\vec{r_2}$ to get $\ket{\alpha_2}$ and $ \ket{\beta_2}$.
    \item Now consider both the impurity together at $\vec{r_1}$ and $\vec{r_2}$. This will give four in-gap modes, $\epsilon_n$, with corresponding wavefunctions, $\ket{\psi_n}$. 
    \item The Hamiltonian of the two impurities in the diagonal basis is: $H(\vec{r_1}, \vec{r_2}) = \sum_{n=1}^4 \epsilon_n \ket{\psi_n}\bra{\psi_n}$. 
    \item Project $H(\vec{r_1}, \vec{r_2})$ in the basis of the independent single impurity problem as: $H_{\text{eff.}}(\vec{r_1}, \vec{r_2}) = \mathcal{U}^{\dagger} H(\vec{r_1}, \vec{r_2}) \mathcal{U}$, where $\mathcal{U}$ is the unitary matrix given by: $\big(
        \ket{\alpha_1}~\ket{\beta_1}~~\ket{\alpha_2}~~\ket{\beta_2}\big)$.
\end{enumerate}
Note that, during the computation, the eigenstates have been fixed to a single gauge choice. The projected Hamiltonian $H_{\text{eff.}}(\vec{r_1}, \vec{r_2})$ is the effective hybridization between two impurities, whose diagonal part gives the onsite term and the off-diagonal term gives tunneling between current loop impurity levels. We find that $H_{\text{eff.}}(\vec{r_1}, \vec{r_2})$ has the form:
\begin{equation}
    H_{\text{eff.}}(r, \theta) = \begin{pmatrix}
        (2-M_{\text{eff}} )\sigma_z & T(r, \theta) \\
        T^{\dagger}(r, \theta) & (2-M_{\text{eff}} )\sigma_z
    \end{pmatrix},
\end{equation}
where $T(r, \theta) = it_x(r, \theta)\sigma_x + it_y(r, \theta)\sigma_y + t_z(r)\sigma_z$ with $r$ and $\theta$ being the relative distance and angle between $\vec{r_1}$ and $\vec{r_2}$. For $M_h=-1$ and $M_d = 1$, we evaluate the $\theta$ and $r$ dependence of the effective hopping in \Fig{smfig8} (a) and (b). While $M_{\text{eff}} \approx 1.32$ the hoping terms are given by, 
\begin{align}
    t_x(r, \theta) & \approx -0.19 \exp(-0.6 r)  \cos \theta \nonumber\\
    t_y(r, \theta) & \approx -0.19 \exp(-0.6 r)  \sin \theta \nonumber\\
    t_z(r) & \approx 0.36 \exp(-0.39 r).
\end{align}
Thus, $T(r, \theta)$ gives the generalized hopping matrix for a QAH insulator model \cite{Agarwala_PRL_2017}.

We also find that as the host goes towards the atomic insulator limit with $M_h \rightarrow -\infty$, the hopping strengths get suppressed as $|t|\sim 0.46(|M_h|)^{-0.99}$ asymptotically, shown in \Fig{smfig8}(c) for $r=2$ and $\theta=0$.

\begin{figure}
    \centering
    \includegraphics[width=1\linewidth]{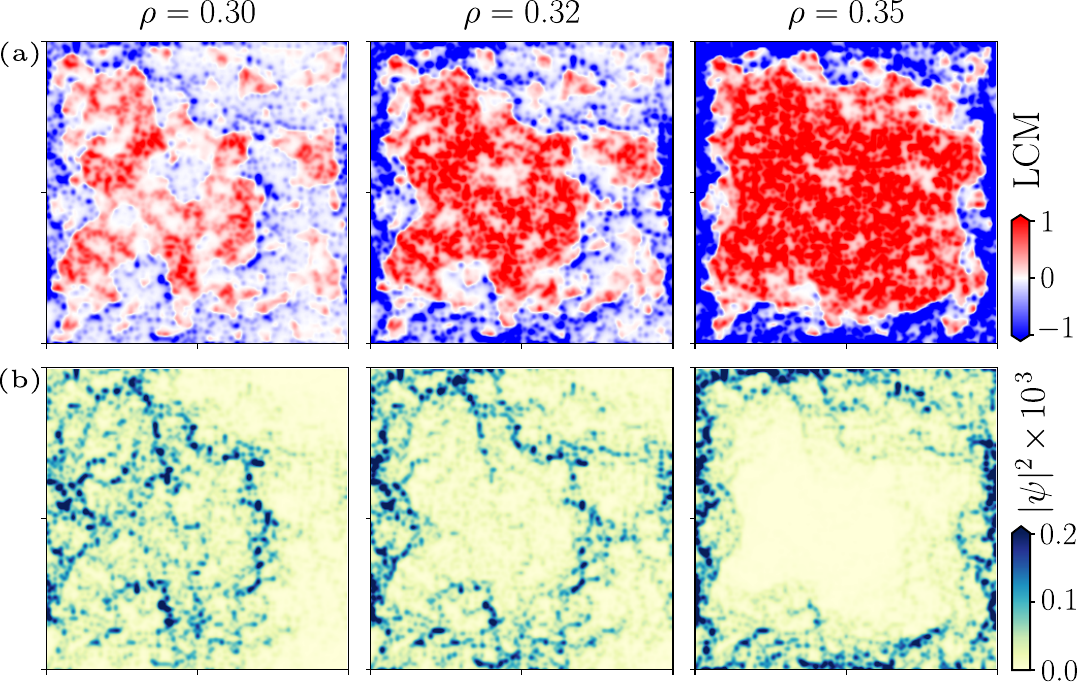}
    \caption{\textbf{Domain percolation:} (a) At $M_h=-1$, snapshots of the local Chern marker (LCM) for three different $\rho$ near the topological transition. (b) Wavefunction amplitudes $|\psi|^2$ of near-zero-energy modes for the same configurations show edge-state formation. The system size is: $100 \times 100$.}
    \label{smfig9}
\end{figure}

\begin{figure}
    \centering
    \includegraphics[width=1\linewidth]{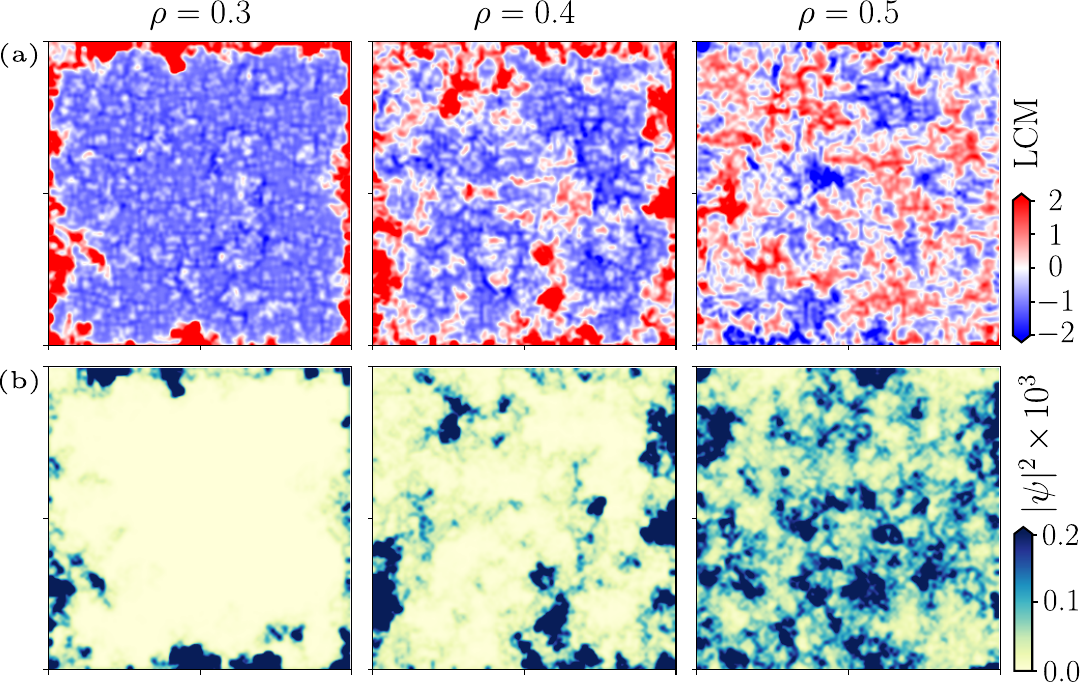}
    \caption{\textbf{Metallic state:} (a) At $M_h=3$, the LCM for three different $\rho$ near the insulator to metal transition. (b) Wavefunction amplitudes $|\psi|^2$ of near-zero-energy modes for the same configurations show the formation of a state extended in the bulk. The system size is: $100 \times 100$.}
    \label{smfig10}
\end{figure}

\section{SVII.~Domain landscape}
Since effective tunneling between impurities resembles hopping in a generalized  Chern insulator, it can give rise to topological domains even at low doping densities. We now discuss how such domains give rise to a topological phase in a trivial host or metallicity in a topological host with opposite chirality.

\subsection{A.~Topological phase transition}
We first consider the negative $M_h$ region of the host insulator. In the main text, domain formation is shown using the ``local Chern marker" (LCM) for $M_h = -1$ and $\rho = 0.28$. Along this $M_h$, the trivial host goes through a topological transition at $\rho \approx 0.308$ (see \Fig{smfig5}(d)). Snapshots of LCM in \Fig{smfig9}(a) show how topological domains (positive LCM) in a trivial host grow and merge with each other with increasing $\rho$ across the transition. The corresponding in-gap states are shown in \Fig{smfig9}(b), which live in the boundaries of the domains (negative LCM) and become a global chiral edge state upon topological transition in the alloy.

\subsection{B.~Edge-state mediated metal}
When the host is also topological with $M_h=3$, we show that domains of both positive and negative LCM appear in the alloy at an intermediate density. This can be seen in \Fig{smfig10}(a) for three different $\rho$. While at $\rho =0.3$, the LCM shows that the bulk of the lattice has a negative value close to $-1$, and the edges have a large positive value. Thus, the alloy is in the QAH phase with the same chirality as the host insulator. As $\rho$ increases, the domains of positive LCM appear on the host, and both type domain coexists. This allows electronic states to proliferate in the bulk via boundaries between domains of opposite chirality. The metallic features in the zero-energy wavefunction can be seen in \Fig{smfig10}(b).

\begin{figure}
    \centering
    \includegraphics[width=0.95\linewidth]{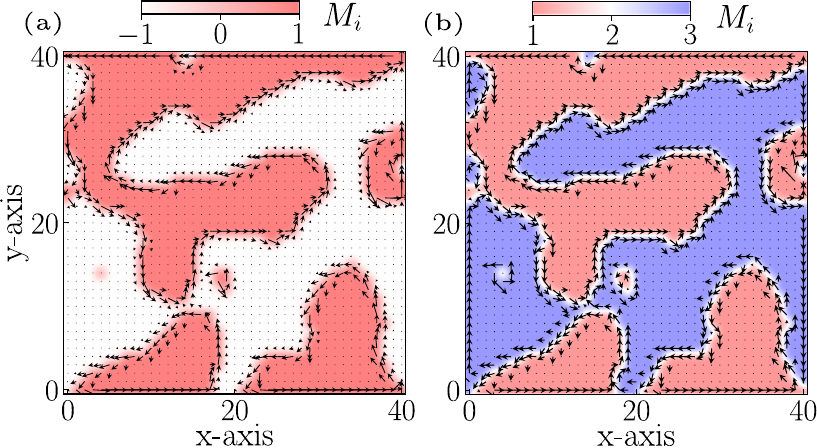}
    \caption{\textbf{Current at domain interface:} (a) Current distribution in $41 \times 41$ lattice with domains of mass term $M_d =1$ (topological with positive LCM) and $M_h=-1$ (trivial with $0$ LCM). (b) Same as (a) but with $M_h = 3$ (negative LCM) such that both domains are topological with opposite chirality. For both snapshots, we use the mass correlation with correlation length $\xi = 2$ to create larger domains.}
    \label{smfig11}
\end{figure}

\subsection{C.~Current profile at domain interface}
To see the current profile in a domain landscape, we brute-force generate a correlation in the mass distribution with correlation length $\xi$ in the alloy as follows:
\begin{equation}
    \langle M_iM_j\rangle \sim \exp \left( -\frac{|r_i - r_j|^2}{2 \xi^2} \right).
\end{equation}
We choose $L=41$ and $\xi = 2$ to calculate the current distribution in the lattice. \Fig{smfig11}(a) shows that when there are only topological domains, like in the case of a topological random alloy with a trivial host, the domain boundaries carry a single edge current. But when the host is topological with opposite chirality, the domain landscape contains two types of topological domains, and at their interfaces, two co-propagating current channels appear as shown in \Fig{smfig11}(b). 

\begin{figure}
    \centering
    \includegraphics[width=1\linewidth]{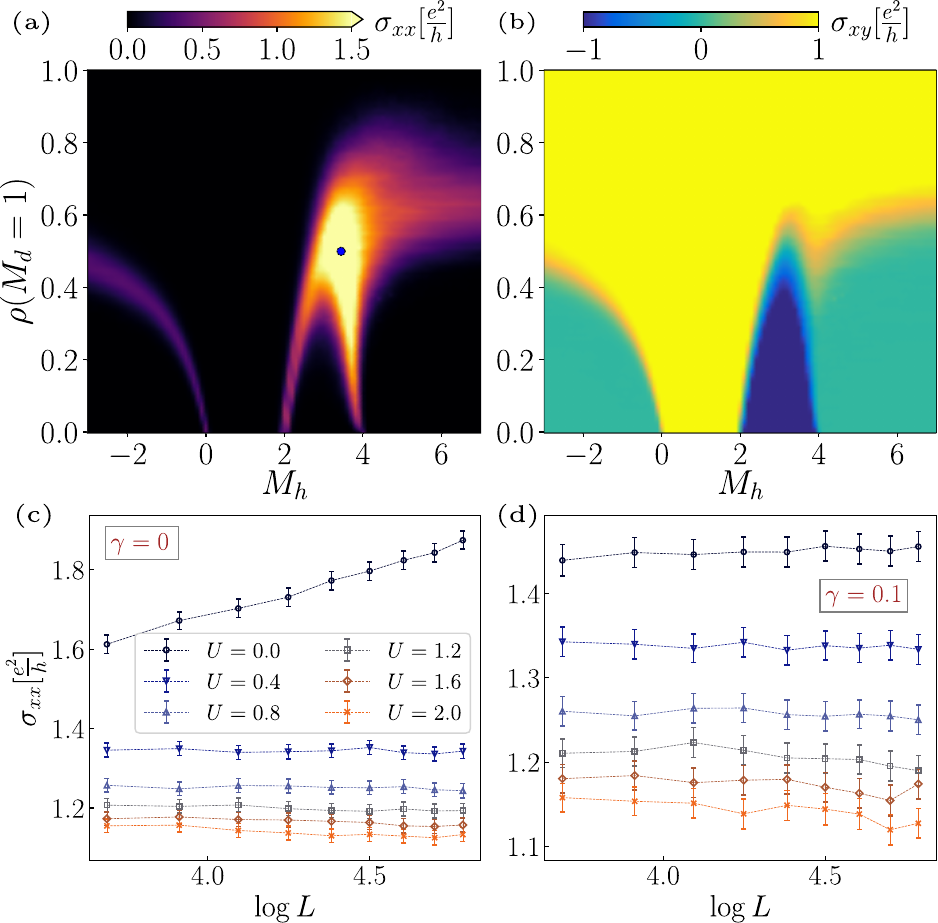}
    \caption{\textbf{Class A alloy:} (a) $\sigma_{xx}$ and (b) $\sigma_{xy}$ phase diagram of the class A topological alloy with particle-hole symmetry breaking term $\gamma = 0.1$ and Anderson disorder of strength $U=1$. For (a) and (b), $30 \times 30$ and $60 \times 30$ systems are used, respectively, and data are averaged over $420$ configurations. (c) Fixing $M_h = 3.5$ and $\rho=0.5$ (marked in (a)), $\sigma_{xx}$ as a function of system size for various $U$ and $\gamma = 0$. (d) Same as (c) but for $\gamma = 0.1$. The geometry used for (c) and (d) is an $L \times 2L$ lattice, and the disorder average is performed over $1400$ configurations.}
    \label{smfig12}
\end{figure}

\begin{figure*}
    \centering
    \includegraphics[width=1\linewidth]{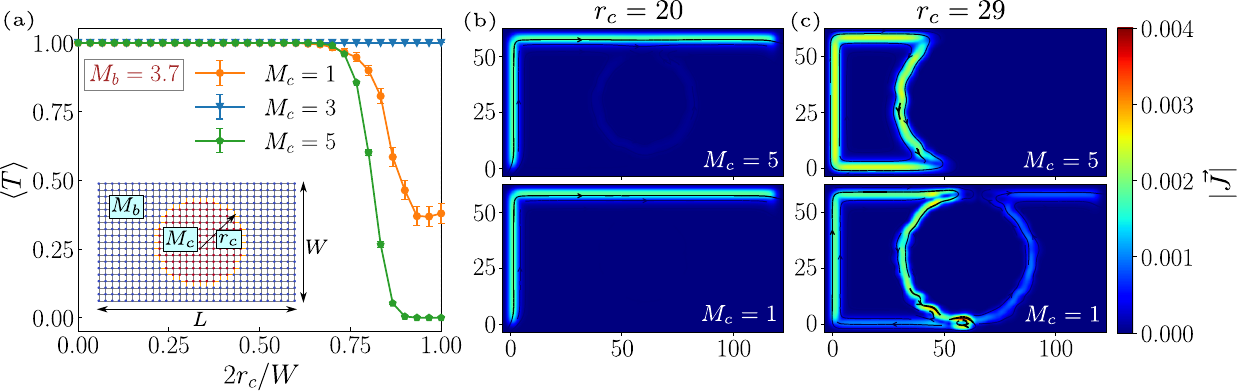}
    \caption{\textbf{Snake state formation:} (a) Configuration-averaged transmission $\langle T \rangle$ from left to right with increasing radius $r_c$ of the central domain in $L \times W$ Chern insulator lattice. The central domain has mass $M_c$, while the background mass is $M_b = 3.7$ ($C = -1$ phase), as shown in the inset. The interface between the domains has a hopping disorder of strength $\delta t = 1$. $L=120$, $W=60$ and the number of configurations used are $400$. (b) Current profile in a single configuration of the same system for $M_c=5, 1$ with $r_c =20$ ($2r_c/W \approx 0.67$) shows edge transport. (c) For $r_c=29$ ($2r_c/W \approx 0.97$) while $M_C = 5$ show reflection $M_c = 1$ shows transmission through domain boundaries.}
    \label{smfig13}
\end{figure*}

\section{SVIII.~Class A topological random alloy}
The class A Hamiltonian for the topological random alloy is given by, 
\begin{align}
	H_{A} & =  \sum_{i, \vec{\eta}}\Big( \Psi^\dagger_{i} \big[T_{\vec{\eta}} + \frac{\gamma}{2}\big] \Psi_{i+\vec{\eta}} + \text{h.c.}  \Big) \nonumber \\
     & ~~~~~~~~~~~~~~~~~~ + \sum_i \Psi^\dagger_i \big[ (2-M_i)\sigma_z + U_i\big] \Psi_i,
	\label{eq_clsAham}
\end{align}
where matrix $T_{\vec{\eta}}$ and the random distribution of $M_i$ for given $\rho$ is described in the main text. The term $\gamma$ explicitly breaks the particle-hole symmetry in the system. We also consider the on-site Anderson disorder $U_i$, drawn from the box distribution $[-U/2, U/2]$. For a large $\gamma$, the clean model without doping will itself have a metallic phase; to avoid that, we consider $\gamma = 0.1$, which allows only insulating phases at $E_F = 0$. At $\rho=0$, the parameter regions of $M_h$ supporting the QAH insulators remain the same as the class D model. With the introduction of doping, the phase diagram of the class A random alloy for disorder strength $U =1$ is shown in \Fig{smfig12}(a) and (b) in terms of configuration-averaged $\sigma_{xx}$ and $\sigma_{xy}$. The qualitative features, such as doping-induced topology below the classical percolation threshold and the emergence of a metal, are robust under class A perturbation. 

In \Fig{smfig12}(c) and (d), we show the variation of configuration-averaged $\sigma_{xx}$ with linear system size $L$ in the metallic phase (marked in \Fig{smfig12}(a)), for $\gamma = 0$ and $0.1$ respectively. With both $\gamma = 0$ and $U = 0$, the system goes back to class D alloy and shows $\sigma_{xx} \propto \log L$ similar to \Fig{smfig3}(b). As we turn on either $U$ or $\gamma$, or both, $\sigma_{xx}$ exhibits the characteristics of a scale-invariant metal for a range of $U$ before the system localizes for large Anderson disorder.

\begin{figure*}
    \centering
    \includegraphics[width=1\linewidth]{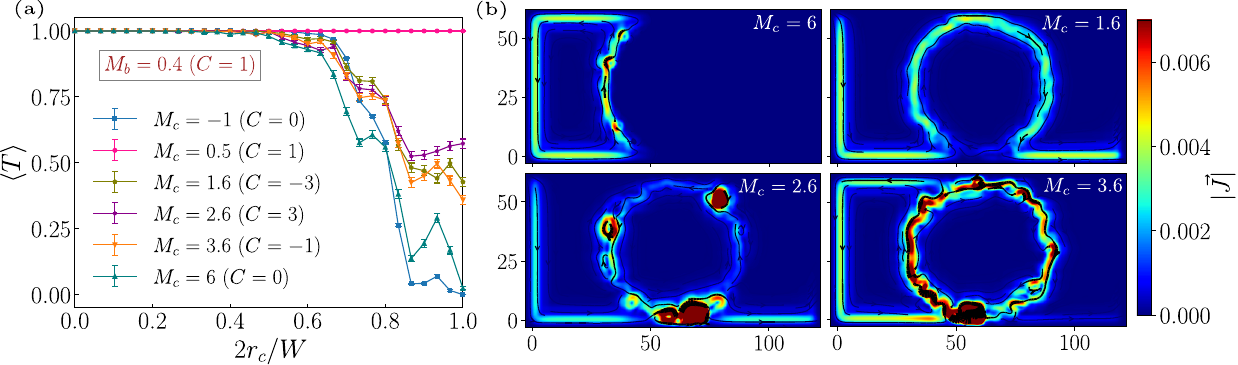}
    \caption{\textbf{High Chern number system:} (a) Transmission (averaged over $400$ configurations) for various $M_c$ with $2r_c/W$ for the high Chern number system. $M_b$ is fixed at $0.4$ with $L=120$ and $W = 60$. (b) For $r_c = 29$, the current profile show reflection for $M_c = 6$ and transmission for $M_c = 1.6, 2.6$ and $3.6$.}
    \label{smfig14}
\end{figure*}

\section{SIX.~Role of chiral edge state}
To understand the role of chiral edge states in forming the metallic phase of the topological random alloy, we construct the following system using the QWZ model. Consider an $L \times W$ lattice, where we create a central domain of radius $r_c$ with mass $M_c$ different from the mass $M_b$ on the remaining sites. We also put disorder in the hopping from box distribution: $[-\delta t/2, \delta t/2]$ with $\delta t = 1$, at the interface between domains to break any accidental crystalline symmetries. The schematic of such a system is shown in the inset of \Fig{smfig13}(a). Using two terminals on the left and right, we evaluate transmission in the system as a function of the $r_c$. In \Fig{smfig13}(a) we show the configuration averaged transmission $\langle T \rangle$ for $M_b=3.7$ (Chern number $C = -1$) and $M_c = 3, 5, 1$. For smaller $r_c$, all three of them have unit quantized $\langle T \rangle$ as the background domain carries one chiral edge state, see \Fig{smfig13}(b). When the central domain is in the same topological phase as in the background for $M_c=3$, the edge state always remains intact, thus $\langle T \rangle = 1$ for all $r_c$. For $M_c = 5$, $\langle T \rangle$ goes to zero around $2r_c/W \sim 0.8$ because the trivial central domain pinches off the edge state of the background, as we gave it a finite localization length in bulk by choosing $M_b=3.7$ ($M_b = 3$, will have the smallest localization length). The pinched-off edge state reflects back from the central domain as shown in (top) \Fig{smfig13}(C). When $M_c = 1$, the central domain is in a $C = 1$ QAH phase hosting an opposite chiral edge state than the background, and $\langle T \rangle$ has a finite signal even in the $2r_c/W > 0.8$ region. If we look at the current profile for this case (see bottom of \Fig{smfig13}(c)), it shows the current flow along the domain boundaries and transmits from left to right. This happens because at the domain boundary two chiral edge states co-propagate, and since the domains have different chiralities, these states hybridize to form a snake state rather than gap out \cite{Lambert_graphenesnake_2008}. This phenomenon seeds the metallic state in the alloys, as shown in \Fig{smfig10} and \Fig{smfig11}.

\subsection{A.~High Chern number QAH system}
We now extend the above idea to higher-Chern-number QAH insulators. Let us consider the following high Chern number (hCN) model on a $L \times W$ lattice \cite{Ji_highCN_2025},
\begin{align}
	H_{\text{hCN}} & = \sum_i \Psi^\dagger_i (2-M_i)\sigma_z \Psi_i \nonumber \\
     & ~-\sum_{i, \vec{\eta}}\Big( \Psi^\dagger_{i} \frac{\sigma_z}{2} \Psi_{i+\vec{\eta}} + \Psi^\dagger_{i} \frac{\sigma_{\eta}}{2} \Psi_{i+\vec{2\eta}} + \text{h.c.} \Big).
	\label{eq_highCham}
\end{align}
With uniform mass $M_i = M \forall i$, the system exhibits the following topologically distinct phases with Chern number, 
\begin{equation}
C =
\begin{cases}
0,  & M < 0, \\
1,  & 0 < M < 1, \\
-3, & 1 < M < 2, \\
3,  & 2 < M < 3, \\
-1, & 3 < M < 4, \\
0, & M > 4.
\end{cases}
\end{equation}
Now fixing the $M_b =0.4$ in the $C= 1$ phase, we show $\langle T \rangle$ as a function of $2r_c / W$ in \Fig{smfig14}(a) for all 6 phases of the central domain. While for smaller $r_c$ all of them show $\langle T \rangle = 1$, which occurs via the edge state similar to the previous case (see \Fig{smfig13}(b)), for the trivial center ($M_c = -1, 6$), the edge state pinches off almost near the geometric limit $2r_c \sim W$. Expectedly, the same topological phase ($M_c = 0.5$) carries on the edge transport with $\langle T \rangle =1$ for all $r_c$. For all three different topological regions in the center, complete pinch off never happens, and $\langle T \rangle$ remains finite. The current profiles shown in \Fig{smfig14}(b) describe the reflection for trivial $M_c= 6$, and transmission for topological $M_c$. Similar to the previous QWZ setup, if the central domains have opposite chirality (i.e, $M_c = 1.6, 3.6$), the transmission occurs via the formation of snake states at the domain boundaries. Hence, we conjecture that a random alloy made out of higher Chern number QAH insulators may exhibit the metallic phase when the dopant and the host material have opposite chiralities. Interestingly, even for $M_c = 2.6$, where the central domain has the same chirality but consists of three edge states, we find finite transmission even after geometric pinch off. This may be a result of partial hybridization of the edge modes. In the future, it will be interesting to see whether such transmission also leads to metallicity in random binary alloys. 
